\documentstyle[epsfig,12pt]{article}
\newcommand{\ba}{\begin{array}}
\newcommand{\ea}{\end{array}}

\newcommand{\be}[1]{
\begin{eqnarray}\label{#1}}
\newcommand{\ee}{\end{eqnarray}}

\newcommand{\insertfig}[2]{\mbox{\epsfxsize=#1cm \epsfbox{#2.eps}}}

\newcommand{\btab}{\begin{tabular}}
\newcommand{\etab}{\end{tabular}}

\newcommand{\ci}[1]{\cite{#1}}

\newcommand{\re}[1]{(\ref{#1})}

\newcommand{\edd}{\end{document}}

\newcommand{\pert}{{\cal D}}
\def\xslash{\rlap/{\mkern-1mu x}}
\newcommand\pbar{\bar{\psi}}
\newcommand\p{\psi }
\def\Gtilde{\tilde{G}}

\newcommand{\alf}{\ifmmode\alpha \else$\alpha \ $\fi}
\newcommand{\bt}{\ifmmode\beta \else$\beta \ $\fi}
\newcommand{\gm}{\ifmmode\gamma \else$\gamma \ $\fi}
\newcommand{\Dl}{\ifmmode\Delta \else$\Delta \ $\fi}
\newcommand{\eps}{\ifmmode\varepsilon \else$\varepsilon \ $\fi}
\newcommand{\dl}{\ifmmode\delta \else$\delta \ $\fi}
\newcommand{\et}{\ifmmode\eta \else$\eta \ $\fi}
\newcommand{\vphi}{\ifmmode\varphi \else$\varphi \ $\fi}
\newcommand{\om}{\ifmmode\omega \else$\omega \ $\fi}
\newcommand{\pl}{\ifmmode\partial \else$\partial \ $\fi}
\newcommand{\ps}{\ifmmode\psi \else$\psi \ $\fi}
\newcommand{\sg}{\ifmmode\sigma \else$\sigma \ $\fi}

\newcommand{\Lam}{\ifmmode\Lambda \else$\Lambda$\fi}

\newcommand{\PRD}[3]{Phys.\ Rev.\ D {\bf {#1}}, {#2} ({#3})}

\newcommand{\NPA}[3]{Nucl.\ Phys.\ A {\bf {#1}}, {#2} ({#3})}
\newcommand{\NPB}[3]{Nucl.\ Phys.\ B {\bf {#1}}, {#2} ({#3})}
\newcommand{\PLB}[3]{Phys.\ Lett.\ B {\bf {#1}}, {#2} ({#3})}

\newcommand{\EPJC}[3]{Eur. Phys. J. \ C {\bf {#1}}, {#2} ({#3})}

\begin{document}
\renewcommand{\thefootnote}{\fnsymbol{footnote}}
\makebox[2cm]{}\\[-1in]
\begin{flushright}
\begin{tabular}{l}
\\
TPR-00-16\\
TUM/T39-00--19\\
\end{tabular}
\end{flushright}
\vskip0.4cm
\renewcommand{\thefootnote}{\fnsymbol{footnote}}
\begin{center}
{\bf\Large 
 Twist-3 contribution to the  
$\gamma^*\gamma\rightarrow \pi\pi$ amplitude \\
in the Wandzura-Wilczek approximation
}\footnote{Work supported in
    part by BMBF}

\vspace{0.5cm}

N. Kivel$^{a,b}$, L. Mankiewicz$^{c,d}$

\vspace{0.5cm}

\begin{center}

{\em$^a$Institut f\"ur Theoretische Physik, Universit\"at Regensburg \\
D-93040 Regensburg, Germany }

{\em $^b$Petersburg Nuclear Physics Institute,
  188350, Gatchina, Russia
}

{\em $^c$ N. Copernicus Astronomical Center, ul. Bartycka 18,
PL--00-716 Warsaw, Poland}

{\em $^d$ Physik-Department, Technische Universit\"{a}t M\"{u}nchen,
D-85747 Garching, Germany}

\end{center}

\centerline{\bf version from  \today}

 \end{center}
 \vspace{1.5cm}
\begin{abstract}
  We have calculated the Wandzura-Wilczek contribution to the twist-3 part of
  $\gamma^*\gamma\rightarrow 2\pi$ amplitude. It describes interaction of the
  longitudinally polarized virtual photon with the real one, and it is
  suppressed by $1/Q$, where $Q^2$ is the virtuality of the $\gamma^*$, as
  compared to the twist-2 contribution. We have found that, in the
  Wandzura-Wilczek approximation, factorization applies to the twist-3
  amplitude.

\end{abstract}

\section{\normalsize \bf Introduction}

Hadron production in the reaction $\gamma^*(q)\gamma(q')\rightarrow {\mbox
  {hadron(s)}}$, $q^2 = - Q^2$, has been a subject of considerable interest in
the QCD community from both experimental \cite{CELLO,CLEO} and theoretical
\cite{Budnev, Lepage, Efremov} points of view.  Recently it has been proposed
\cite{Dittes,Diehl} to investigate a process $\gamma^*\gamma\rightarrow\pi\pi$
when the two pion state has a small invariant mass. Thanks to the QCD
factorization theorem \cite{Freund}, in the leading twist approximation the
amplitude can be represented as a convolution of perturbatively calculable
Wilson coefficients and new non-perturbative objects, the so-called two-pion
distribution amplitudes ($2\pi$DA's). They are given by matrix elements of
twist-2 QCD string operators between vacuum and the two-pion state \cite{Diehl,
  Polyakov}.  Moreover, $2\pi$DA's can be related by the crossing symmetry to
skewed parton distributions \cite{Ji97,Rad97} which recently have been subject
of considerable interest. Later, NLO corrections to the leading twist amplitude
have been calculated \cite{KMP} and the recent thorough phenomenological
analysis \cite{Diehl1} have shown that experimental studies of 2$\pi$
production cross-section are possible with existing $e^+e^-$ facilities.

In general, parity invariance restricts the number of independend helicity
amplitudes in the $\gamma^*\gamma\rightarrow\pi\pi$ process to three, which
correspond to different possible projections of the total angular momentum on
the $\gamma^* \gamma$ collision axis, $J_z = 0,\pm 1,\pm 2$. $J_z = 0$ and $J_z
= \pm 2$ amplitudes scale as $Q^2 \to \infty$, and their corresponding QCD
representation has been discussed in \cite{Diehl,Polyakov,KMP}. The amplitude
corresponding to $J_z = \pm 1$ is suppressed as $1/Q$ and has not been
discussed so far. On the other hand, as it has been demonstrated in
\cite{Diehl1}, it can be extracted from the experimental data by a suitable
weighting procedure.

The purpose of this paper is to fill this gap in the QCD description of the
$\gamma^*\gamma\rightarrow\pi\pi$ amplitude. As it can be anticipated from its
$1/Q$ suppression, the leading contribution to the $J_z = \pm 1$ amplitude is
given by matrix elements of twist-3 quark-quark and quark-gluon operators. In
this paper we restrict ourselves to the Wandzura-Wilczek (WW) approximation
\cite{Wandzura} i.e., we consider only the contribution from quark operators.
Recall that the recent experimental analysis \cite{g2} of the polarized nucleon
structure function $g_T(x,Q^2)$ indicates that the WW approximation can account
rather well for the experimental data.  A similar conclusion is reached in the
instanton model of the QCD vacuum \cite{instanton}, where quark-gluon
correlations in a nucleon are suppressed by a small parameter given by the
packing fraction of instantons. Our results can be therefore considered as an
estimate of the order of magnitude of the $J_z = \pm 1$ amplitude.

The problem of twist-3 contributions to the hard exclusive amplitudes has
recently acquired a considerable attention
\cite{Anikin,Penttinen,BM,KPST,RadWeiss,BlRo}. In particular, the approach 
presented here is rather similar in spirit to the analysis of \cite{RadWeiss},
which presented a detailed consideration of the problem
of the DVCS amplitude on a pion.

\section{\normalsize \bf General definitions}

Kinematics of the reaction $\gamma^*(q) \gamma(q') \to \pi(k_1) \pi(k_2)$
can conveniently  be described in terms of a pair of light-like vectors $p,\ z$
which obey
\be{nvectors}
\ba{l}
p^2 = z^{2} = 0, \, p \cdot z \not = 0
\ea
\ee
and define longitudinal directions. Here $p \cdot z = p_\mu z^\mu$. 
Let $P$ and $k$ denote total and relative
momenta of the $\pi$ meson pair, respectively,
\be{pivectors}
\ba{l}
P^2=(k_1+k_2)^2=W^2, k^2=(k_1-k_2)^2=4 \, m_\pi^2 - W^2, P\cdot k = 0 
\ea
\ee

The initial and final states momenta can be decomposed as

\be{lcexpansion}
\ba{l}
\displaystyle
q = p  - \frac{Q^2}{2 (p\cdot z)}z, \quad q^2 = - Q^2\, 
\hspace*{8mm}
q^\prime = \frac{Q^2+W^2}{2 (p\cdot z)}\, z, \quad q^{\prime \, 2} = 0
\\[4mm]\displaystyle
P = q + q^\prime = p + \frac{W^2}{2 (p\cdot z)} \, z, \quad  P^2 = W^2
\\[4mm] \displaystyle
k = \xi p - \frac{\xi W^2}{2 (p\cdot z)} \, z  + k_\perp
\ea
\ee
The longitudinal momentum distribution between pions is described by the
variable $\xi = (k\cdot z)/(p\cdot z)$. Alternatively,
$$
\xi = \beta \cos\theta_{\rm cm}\, ,
$$
where $\theta_{\rm cm}$ is the polar
angle of the pion momentum in the CM frame with respect to
the direction of the total momentum $P$ and $\beta$ is the velocity of
produced pions in the center-of-mass frame
\begin{eqnarray}
\nonumber
\beta=\sqrt{1-\frac{4 m_\pi^2}{W^2}}\, .
\end{eqnarray}

The amplitude of
hard photo-production of two pions
is defined by the following matrix element
between vacuum and two pions state:
\be{defT}
T^{\mu\nu} = i \int d^4x e^{-ix\cdot {\bar q}}
\langle 2 \pi (P, k)| T J^\mu(x/2) J^\nu(-x/2) | 0 \rangle\, ,
\mskip10mu {\bar q}=\frac12(q-q')\, 
\ee
where $J^\mu(x)$ denotes quark electromagnetic current.  Hard photo-production
corresponds to the limit $Q^2 \gg \, W^2 \geq \Lambda_{\rm QCD}^2$ where the
amplitude (\ref{defT}) can be represented as an expansion in terms of powers
of $1/Q$. According to the factorization theorem the leading twist term in the
expansion can be written as a convolution of hard and soft blocks. The
coefficient functions can be calculated from appropriate partonic subprocesses
$\gamma^*+\gamma\rightarrow \bar{q}+q$ or $\gamma^*+\gamma\rightarrow g+g$.
Soft blocks, describing transition of partons into the final meson pair are
given by quark and gluon $2\pi$ distribution amplitudes.

According to the analysis of Ref. \cite{KMP} the 
amplitude $T^{\mu\nu}$ is a sum of three terms
\be{Tdecompos}
T^{\mu\nu}(q,q',P,k)& = & \frac i2(-g^{\mu\nu})_T T^{\gamma\pi\pi}_0(q,q',P,k)+
i\frac1{Q^2}k_\perp^\nu (P+q')^\mu T^{\gamma\pi\pi}_1(q,q',P,k)+ 
\nonumber \\
&+&\frac i2\frac{k_\perp^{(\mu} k_\perp^{\nu)}}{W^2} 
\, T^{\gamma\pi\pi}_2(q,q',P,k)
\ee
where $(-g^{\mu\nu})_T=\left( \frac{p^\mu z^{\nu} + p^\nu z^{\mu}}{p \cdot z}
  - g^{\mu\nu}\right)$ is the metric tensor in the transverse space and
$k_\perp^{(\mu} k_\perp^{\nu)}$ denotes traceless, symmetric tensor product of
relative transverse momenta \re{lcexpansion}.

Amplitudes $T^{\gamma\pi\pi}_0$ and $T^{\gamma\pi\pi}_2$ correspond to $J_z =
0$ and $J_z = \pm 2$, respectively. They have been discussed
at length in \ci{Diehl,Freund,Polyakov,KMP}
In this paper we consider the $J_z = \pm 1$ amplitude $T^{\gamma\pi\pi}_1$. It
arises when 
the (real) transverse photon collides with the (virtual) longitudinal one and
produces a pair of pions in $L_z = \pm 1$ state. It is easy to see that such
a process requires helicity flip along the quark line, and hence it vanishes
in the leading twist approximation.

Consider now a frame in which the pion pair is moving in the positive ${\hat
  z}$ direction so that $p^+$ and $z^-$ are the only nonzero components of $p$
and $z$, respectively.  In an infinite momentum frame $p^+ \sim Q \rightarrow
\infty$ with fixed $(p\cdot z)\sim 1$. From \re{lcexpansion} it follows that
in this frame $(k\cdot z)\sim 1$ and $k_\perp \sim Q^0$. Hence, the
contribution due to $T^{\gamma\pi\pi}_1$ is suppressed by $1/Q$ as compared to
the leading one.

In this paper we will consider quark $2\pi$DA's, defined as matrix elements of
the light-cone quark string operator:
\be{quarkDA}
\langle \pi\pi(P, k) | \frac{1}{N_f}\sum_q {\bar q}(z)[z,-z] {\hat z} q(-z) 
|0\rangle
& = & (p\cdot z) \int_0^1 du \, \Phi^Q(u,\xi,W^2) \, e^{i(2u-1)(p\cdot z)} \,
,  
\nonumber \\
&=& (p\cdot z) \int_{-1}^1 dv \, H^Q(v,\xi,W^2) \, e^{i v (p\cdot z)} \, ,
\ee
with ${\hat x} = \gamma^\mu x_\mu$ and
\be{HtoPhi}
H^Q(v,\xi,W^2) = \frac{1}{2}\Phi^Q(\frac{1+v}{2},\xi,W^2) \, .
\ee
$[x,y]$ denotes a path-ordered exponential 
$[x,y] =\mbox{\rm Pexp}[ig\!\!
\int_0^1\!\! dt\,(x-y)_\mu A^\mu(tx+(1-t)y)] 
$ 
which ensures gauge-invariance of the matrix element. In \re{quarkDA} we
have introduced both types of single-varaiable distributions which can be
found in the literature - the distribution amplitude $\Phi^Q(u,\xi,W^2)$ and
the symmetric form $H^Q(u,\xi,W^2)$. As we shall see later,
calculations in the coordinate-space naturally require to introduce a
parametrisation in terms of a double-distribution function introduced in
\ci{Rad}:
\be{DD}
\langle 2\pi(P,k)|\pbar (x)[x,-x]{\hat x} \p (-x)|0\rangle &=&
(P \cdot x)\int d[\alf,\bt] F(\alf,\bt,W^2)e^{i\alf(P \cdot x)+i\bt(k\cdot x)}+
\nonumber \\
&&+(k\cdot x)\int_{-1}^{1}d\bt D(\bt,W^2)e^{i\bt(k\cdot x)}+ O(x^2),
\ee
where
\be{mera}
\int d[\alf,\bt]\equiv 
\int_{-1}^{1} d\alf \int_{-(1-|\alf|)}^{(1-|\alf|)}d\bt \, .
\ee
The second term in \re{DD} corresponds to the so-called ``D-term'' introduced
in \ci{PW}.  Comparing matrix element \re{DD} evaluated on the light-cone,
$x^2 = 0$, with parametrisation \re{quarkDA} one easily obtains a relation
between double and $2\pi$ distribution (\ref{HtoPhi}):
\be{DDtoDA}
H^Q(u,\xi,W^2)= \int d[\alf,\bt]F(\alf,\bt,W^2)\delta(\alf+\xi\bt-u)+ 
\theta(\vert u \vert < \vert \xi \vert )\, sign(\xi)\, D(u/\xi) \, .
\ee
Note that the matrix element of the pseudovector quark string operator 
\be{zero}
\langle \pi\pi(P, k) | \frac{1}{N_f}\sum_q {\bar q}(z)[z,-z] {\hat z} \gamma_5
q(-z) |0\rangle
= 0 
\ee
vanishes because of the positive parity of the final pion pair.

Finally, we recall that all distributions introduced here depend also on a
normalization scale $\mu$. This dependence is not relevant for our discussion
and has been neglected. For a sake of simplicity from now on we will also
neglect the gauge factors $[x,y]$ and, wherever possible, the explicit $W$ -
dependence.

\section{\normalsize \bf Angular momentum analysis of $\gamma^*
  \gamma \to \pi \pi$ amplitude}

Our goal here is to identify the leading contribution to the amplitude
$T^{\gamma\pi\pi}_1$. For simplicity
we will consider the massless quark degrees of freedom only. Taking into
account quark masses and gluonic degrees of freedom does not change our
conclusions qualitatively. A convenient coordinate frame is obtained by
identifying the null vectors $p$ and $z$ introduced in 
\re{lcexpansion} with a pair $n,n^*$ which obeys
\be{nvectors1}
\ba{l}
n^2 = n^{*2} = 0, \, n \cdot n^* = 2,\\
n = (1,0,0,-1), \, n^* = (1,0,0,1)\, .
\ea
\ee
Taking
\be{pzvectors}
\ba{l}
z = n, \, p = \frac{Q}{2}\ n^* \ .
\ea
\ee
one obtains a collinear collision of $\gamma^*$ and $\gamma$, with the final
pion pair moving in the virtual photon direction.  The virtual photon can
carry both transverse and longitudinal polarizations.  For asymptotically
large $Q^2$ the interaction between photons occurs at a light-like separation.

Let us consider first the tree-level contribution.  Corresponding space-time
configuration is depicted on Fig.\ref{SpaceTime0}. The real photon and the
final quark-antiquark pair move along light rays defined by $x_{-}= x_0 - x_3 =
0$ and $x_+ = x_0 + x_3 = 0$, respectively. If both photons have the same
helicities, the projection of the angular momentum onto the collision axis $J_z
= 0$. Vector couplings create the $q {\bar q}$ pair with opposite
polarizations, and the angular momentum conservation requires that the pair has
the orbital angular momentum $L_z = 0$.  Transition of such a pair into a 2
pions with $L_z = 0$ is described by twist-2 matrix element \re{quarkDA}.

Longitudinal polarization of the virtual photon leads to a configuration with
$J_z = \pm 1$, see Fig.\ref{SpaceTime1}. Due to the angular momentum
conservation, the quark-antiquark pair is created with the intrinsic orbital
angular momentum $L_z = J_z$. Subsequently, it evolves into a pion pair with
$L_z = \pm 1$. The amplitude must be therefore proportional to $k_\perp$. On
the other hand, non-zero quark orbital angular momentum requires that the
transition occurs through matrix element of a quark twist-3 operator, see the
next section, which results in a suppression $k_\perp/Q$ as compared with the
amplitude corresponding to $J_z=0$.

So far we have considered only a point-like interaction of the real photon with
the virtual one. Another contribution arises when the real photon splits into
its quark-antiquark component long before its interaction with $\gamma^*$.
Configuration arising if the transition is described by twist-2, $L_z = 0$
photon distribution amplitude is shown in Fig.\ref{SpaceTime2}. From
dimensional considerations it follows that twist-2 matrix element between a
real photon and the vacuum is parametrized by a constant carrying dimension
one. It is usually defined as $f_\gamma = \chi \langle {\bar \psi} \psi
\rangle$, where $\chi$ is the vacuum magnetic susceptibility, and $\langle
{\bar \psi} \psi \rangle$ is the quark condensate \cite{photon}. Such a
contribution would therefore provide a term in $T^{\gamma\pi\pi}_1$ of the
order of $f_\gamma/Q$.  Note, however, that the angular momentum conservation
requires that the $q {\bar q}$ pair has to acquire in the hard collision two
units of orbital angular momentum. It means that twist-2 contribution to this
amplitude vanishes to all orders in perturbation theory.  The situation here is
similar to the computation of the amplitude of hard photoproduction of
transverse vector mesons by longitudinal photon, discussed in
\cite{MPW,CollinsDiehl}. Possible higher-twist corrections to this mechanism
will be suppressed by at least one power of $Q$. We conclude that the process
described in the previous paragraph provides the leading contribution to
$T^{\gamma\pi\pi}_1$.

\section{\normalsize \bf Twist-3 amplitude $T^{\gamma\pi\pi}_1$ in the
Wandzura-Wilczek approximation}

In this section we employ technique developed in \ci{BB89} to obtain
twist-3 contribution to the amplitude $T^{\gamma\pi\pi}_1$ in the
Wandzura-Wilczek approximation i.e., neglecting all quark-gluon operators. Our
stating object is the amplitude \re{defT} written in the form:
\be{T}
T^{\mu\nu} = 16i \int d^4x e^{-ix(q-q')}
\langle 2 \pi (P, k)| T J^\mu(x) J^\nu(-x) | 0 \rangle\, .
\ee
The main contribution to the above integral arises from distances in the
vincinity of the light cone,  $x^2\sim 1/Q^2$. 

\begin{figure}[htb]
\begin{center}
\hspace{0cm}
\insertfig{8}{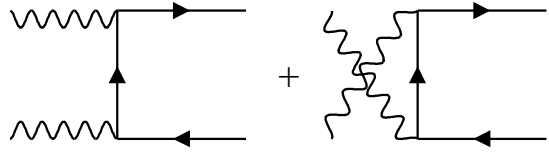}
\end{center}
\vspace{-0.5cm}
\caption[dummy]{
\small 
Leading contribution to the amplitude \re{defT} in the scaling limit.
\label{fig4}
}
\end{figure}

Let us first consider a situation where both photons 
are virtual and their momenta are spacelike:
\be{mom}
-q^2=Q^2\gg \Lambda_{QCD}, -q'^2=Q'^2\gg  \Lambda_{QCD}
\ee
To extract the part of the amplitude which corresponds to the case when the one
of the two photons is longitudial we have to separate contribution which is
proportional to transverse momentum $k_\perp$.  An analogous problem in DVCS
on a pion has been considered recently in
\ci{Anikin,Penttinen,BM,KPST} and \cite{RadWeiss}.  Here, we present a
solution obtained using coordinate-space technique which is close
to 
the approach of \cite{RadWeiss}.
 
Let us consider the $T$-product of two electromagnetic currents. In the
vincinity of the light cone $x^2 \approx 0$ the dominant terms are given by
diagrams depicted on Fig.~\re{fig4}. Their contribution can be written as:
\be{1th}
 T\{ J^\mu(x) J^\nu(-x)\}&=&\frac{1}{16\pi^2}\frac1{x^4}\biggl\{
t^{\mu \nu}_{\lambda \sigma}{x_\lambda}
[\pbar (x)\gamma_\sigma \p (-x)- \pbar (-x)\gamma_\sigma \p (x)]+
 \nonumber \\
&&+i\epsilon_{\mu \nu \lambda \sigma }{x_\lambda}
[\pbar (x)\gamma_\sigma\gamma_5  \p (-x)+
\pbar (-x)\gamma_\sigma\gamma_5  \p (x)]\biggl\} \, .
\ee
Here we have introduced a shorthand notation:
\be{tmunu}
t^{\mu \nu}_{\lambda \sigma}=g_{\mu \lambda}g_{\nu\sigma}+
g_{\mu \sigma}g_{\nu\lambda}-g_{\mu \nu}g_{\lambda \sigma} \, .
\ee
Our goal now is to determine matrix elements of vector and pseudovector
non-local operators $\pbar(x)\gamma_\sigma \p(-x)$ and $\pbar(x)\gamma_\sigma
\gamma_5 \p(-x)$.
To this end we employ the operator identity derived in \ci{BM, KPST,RadWeiss}:
\be{Oident2}
\pbar ( x)\gamma_\mu \p (- x)&=&
\frac{\pert_\mu}{(\pert x)}\pbar ( x)\xslash \p (- x)+
\nonumber \\[4mm]&&\mskip-20mu
\frac12\int_0^1 d\alpha 
\left\{
e^{\bar \alpha (x\pert)}+e^{-\bar \alpha (x\pert)}
\right\}
\left[
\partial_\mu -\frac{\pert_\mu}{(x\pert)} (x\partial)
\right]
\pbar (\alpha x)\xslash \p (-\alpha x)+
\nonumber \\[4mm]&&\mskip-20mu
\frac12\int_0^1 d\alpha 
\left\{
e^{-\bar \alpha (x\pert)} - e^{\bar \alpha (x\pert)}
\right\}
i\epsilon_{\mu ijk}x_i\frac{\pert_j}{(\pert x)}
\partial_k \pbar (\alpha x)\xslash\gamma_5 \p (-\alpha x)+\dots \nonumber \\
\ee
where $\bar\alpha=1-\alpha$, $\partial_\mu = \partial /\partial x_\mu$,
and ellipses denote contributions of three-point quark-gluon operators which
are neglected in the WW approximation.  Here, $\pert_\mu$ is a derivative with
respect to translation introduced in \ci{BB89}:
\be{Dtot}
\pert_{\alpha}\left\{ \pbar(tx)
\Gamma [tx, -tx] \p(-tx) \right\} \equiv
\left. \frac{\partial}{\partial y^{\alpha}}
\left\{ \pbar(tx + y) \Gamma
[tx + y, -tx + y] \p(-tx + y)\right\} \right|_{y \rightarrow 0},
\ee
with a generic Dirac matrix structure $\Gamma$. Hence, when acting on a matrix
element between vacuum and the final state with a four-momentum $P_\mu$,
$\pert_\mu \to i P_\mu$.  Identity \re{Oident2} allows to compute matrix
element of $\pbar (x)\gamma_\sigma \p (-x)$ in terms of the matrix element of
the symmetric operator $\pbar (x) {\hat x} \p (-x)$, given by \re{DD}.  Taking
matrix elements of both sides of (\ref{Oident2}) between 2-pion state and the
vacuum one obtains after simple manipulations:
\be{me}
\langle 2\pi(P,k)|\pbar (x)\gamma_\sigma \p (-x)|0\rangle &=&
P_\sigma F_1(Px, kx)+ k_\sigma D_1(Px, kx)+
\nonumber \\
&& i [k_\sigma(Px)-P_\sigma(kx)] [ F_2(Px, kx)+D_2(Px, kx) ] .
\ee
Here we have introduced a compact notation
\be{Fi}
F_1(Px, kx)&=&\int d[\alf,\bt]F(\alf,\bt)e^{i\alf(Px)+i\bt(kx)},
\nonumber \\
F_2(Px, kx)&=&\frac12\int d[\alf,\bt]F(\alf,\bt)\bt\, 
\int_0^1 dt\, t[ e^{i\bar t(Px) } + e^{-i\bar t(Px)} ]
e^{it[\alf(Px)+i\bt(kx)]},
\nonumber \\
D_1( kx)&=&\int_{-1}^1 d\bt D(\bt) e^{i\bt (kx)},
\nonumber \\
D_2(Px, kx)&=&\frac12 \int_{-1}^1 d\bt D(\bt) 
\int_0^1 dt\, t[ e^{i\bar t(Px) } - e^{-i\bar t(Px)} ]
e^{i \bt t (kx)} \, .
\ee

Consider now the pseudovector operator $\pbar (x)\gamma_\mu\gamma_5 \p (-x)$.
Using the identity \ci{BB89}:
\be{avect}
\mskip-10mu
\pbar (x)\gamma_\mu\gamma_5 \p (-x) &=&\int_0^1dt
{ \partial \over \partial x_\mu}\,\pbar(tx)\xslash \gamma_5
\p (-tx)- i\epsilon_{\mu\nu\alpha\beta}\int_0^1dtt\,x^\nu\pert^\alpha
\left[ \pbar(tx)\gamma^\beta  \p (-tx)\right]
\nonumber\\
& &\mskip-50mu{}- \int_0^1\,dt\,\int_{-t}^t\,dv\,\pbar(tx)
\left\{
t\, g\Gtilde_{\mu\nu}(vx)+ v\,\gamma_5 igG_{\mu\nu}(vx)
\right\}
x^\nu \xslash \p (-tx) \, , \nonumber\\
\ee
one can express its matrix elements in the WW approximation through matrix
elements of the vector operator $\pbar \gamma_\sigma \p$. In this way one
obtains 
\be{meps}
\langle 2\pi(P,k)|\pbar (x)\gamma_\sigma \gamma_5 \p (-x)|0\rangle =
\epsilon_{\sigma \lambda \rho \delta } x^\lambda P^\rho k^\delta [F_3(Px,kx) +
D_3(Px,kx)]
\ee
where
\be{F3}
F_3(Px, kx)&=&\frac12\int d[\alf,\bt]F(\alf,\bt)\bt\, 
\int_0^1 dt\, t[ e^{i\bar t(Px) }- e^{-i\bar t(Px)} ]
e^{it[\alf(Px)+i\bt(kx)]},
\nonumber \\
D_3(Px, kx)&=&\frac12 \int_{-1}^1 d\bt D(\bt) 
\int_0^1 dt\, t[ e^{i\bar t(Px) }+e^{-i\bar t(Px)} ]
e^{i \bt t (kx)} \, .
\ee

Now, combining \re{me} and \re{meps} with \re{1th} and \re{T} one finally finds
\be{Tin}
T^{\mu\nu}& =& \frac{i}{\pi^2} \int d^4x e^{-ix(q-q')}
\frac{1}{x^4}\biggl\{ 
t^{\mu\nu}_{\lambda\sigma}x_\lambda P_\sigma \{F_1(Px, kx)-\bar F_1(Px, kx)\}+
\nonumber \\
&+&i x_\mu x_\sigma[k_\perp^{\nu},P^{\sigma}]
\{F_2(Px, kx)-F_3(Px, kx)+ \bar F_2(Px, kx)+\bar F_3(Px, kx)  \}+
\nonumber \\
&+& i x_\nu x_\sigma[k_\perp^{\mu},P^{\sigma}] 
\{F_2(Px, kx)+F_3(Px, kx)+ \bar F_2(Px, kx)- \bar F_3(Px, kx)\}+
\nonumber \\
&+& i x^2 [k_\perp^{\mu},P^{\nu}]\{\bar F_3(Px, kx)- F_3(Px, kx)\}+ \cdots
\biggl\}\, ,
\ee
where for simplicty we have denoted by ellipses all D-type contributions,
$\bar F_i(Px, kx)= F_i(-Px, -kx)$ and
\be{kperp}
[k_\perp^{\mu},P^{\sigma}]= k^\mu_\perp P^\nu-
 k^\nu_\perp P^\mu=k^\mu P^\nu- k^\nu P^\mu +O(W^2/Q^2) \, . 
\ee
The last line follows from the light-cone
expansion \re{lcexpansion}. 

Calculation of the Fourier integrals is now straightforward. Note that at this
point the advantage of using the double-distribution representation for the
matrix element \re{DD} becomes clear, as all dependence on the momentum $k^\mu$
is 
in the exponents. Separating the linear in $k_\perp$ term one obtains
after simple manipulations:
\be{Tout}
T^{\mu\nu}_{1F} &=& i\sum e_q^2\frac1{Q^2} \int d[\alf,\bt] F(\alf,\bt) 
\biggl[ k^\nu_\perp(P+\omega q')^\mu 
 \partial_\xi C(\alf+\xi\bt ,w)+
\nonumber \\
&+& k^\mu_\perp(P-\omega q)^\nu \partial_\xi C(\alf+\xi\bt,-w) 
\biggl] \, ,
\ee
where
\be{omega}
\omega=\frac{q^2-q'^2}{q^2+q'^2},\quad   -1\leq \omega\leq 1,
\ee
and
\be{C}
C(u,\omega)=\frac2{\omega^2(1-u)}
\ln \biggl(\frac{1+\omega u}{1+\omega}\biggl)- 
\frac2{\omega^2(1+u)}
\ln \biggl(\frac{1-\omega u}{1+\omega} \biggl)\, ,
\ee
Using an identity:
\be{red}
\int d[\alf,\bt] F(\alf,\bt) \partial_\xi C(\alf+\xi\bt ,w)=
\partial_\xi\int_{-\infty}^{+\infty}du \int d[\alf,\bt] F(\alf,\bt)
\delta(\alf+\xi\bt-u)C(u,w)=
\nonumber \\
\partial_\xi\int_{-1}^{1}du H_F^Q(u,\xi)C(u,w),\,  \mbox{ where }\, 
H_F^Q(u,\xi)=\int d[\alf,\bt] F(\alf,\bt)\delta(\alf+\xi\bt-u)\, , 
\mskip30mu 
\ee
one arrives at
\be{T1F}
T^{\mu\nu}_{1F}=i\sum e_q^2\frac1{Q^2}
\int_{-1}^{1}du\partial_\xi H_F^Q(u,\xi)
\biggl[ k^\nu_\perp(P+\omega q')^\mu  C(u ,w)+
k^\mu_\perp(P - \omega q )^\nu C(u,-w) .
\biggl]
\ee
Calculation of terms involving D-functions is very similar. We have 
checked that all the D-type contributions can be incorporated
together with F-type terms into 
\be{T1}
T^{\mu\nu}_{1}=i\sum e_q^2\frac1{Q^2}
\int_{-1}^{1}du\, \partial_\xi H^Q(u,\xi)
\biggl[ k^\nu_\perp(P+\omega q')^\mu  C(u ,w)+
k^\mu_\perp(P - \omega q )^\nu C(u,-w) 
\biggl]
\ee  
with the coefficient function $C(u ,w)$ given in \re{C} and the $2\pi$
distribution function
$H^Q(u, \xi)$ defined according to \re{DDtoDA}.

The coefficient function $C(u ,w)$ has a smooth behaviour in the limit
$w\rightarrow 0$:
\be{w0}
C(u ,w)=2u+O(w),
\ee
In the limit when photon with momentum $q'$ becomes real, $q'^2 \to 0$ or
equivalently $w\rightarrow 1$ one finds:
\be{w1}
C(u ,w=1)=\frac2{(1-u)}
\ln \biggl(1-\frac{1- u}{2}\biggl)- 
\frac2{(1+u)}
\ln \biggl(1-\frac{1+u}{2} \biggl)\, , 
\ee  
but
\be{w1s}
\lim_{w\rightarrow 1} C(u,-w)=\ln (1-w)\frac{-4u}{(1+u)^2}+O(1+w) \, ,
\ee
such that the amplitude \re{T1} is singular in this limit. This singularity
cancels, however, when contribution to the scattering amplitude is evaluated
by contraction with polarization vectors of photons with helicities $\lambda$
and $\lambda'$, respectively:
\be{scat}
T_1(\lambda,\lambda') = T_1^{\mu\nu} e_\mu(\lambda) e_\nu(\lambda') \, .
\ee
The second term in \re{T1} can give a non-zero contribution only when a
transverse photon with momentum $q$ collides with a longitudinal photon with
momentum $q'$. All other helicity combinations lead to vanishing
contribution due to contraction of transverse polarisation
vectors with longitudinal momenta.  Note that in the c.m.s. of two photons
the logitudinal polarisation vector $e^\nu(0)$ can be written as \ci{Budnev}:
\be{e0}
e_\nu(0)=i\left(\frac{-q'^2}{(qq')^2-q^2q'^2}\right)^{1/2}\biggl(q-q'
\frac{(qq')}{q^2}\biggl)_\nu \, .
\ee 
Evaluating the contraction \re{scat} one finds that an additional factor
$\sqrt{1-w}$ arises which cancels the logarithmic singularity \re{w1s}.
Hence, for the amplitude $T_1^{\gamma\pi\pi}$ of a collision of the virtual
photon with momentum $q$ with a real photon with momentum $q'$, as introduced
in \re{Tdecompos}, one finds:
\be{T1real}
T_1^{\gamma\pi\pi}=\sum e_q^2 \ \int_{-1}^{1}du\, \partial_\xi H^Q(u,\xi)
\biggl[\frac2{(1-u)}
\ln \biggl(1-\frac{1- u}{2}\biggl)- 
\frac2{(1+u)}
\ln \biggl(1-\frac{1+u}{2} \biggl)
\biggl]\, ,
\ee  
or in terms of $2\pi$ distribution amplitude 
\be{T1fin}
T_1^{\gamma\pi\pi}=\sum e_q^2 \ \int_{0}^{1}du\, \partial_\xi \Phi^Q(u,\xi)
\biggl[\frac{\ln (1-u)}{u}- 
\frac{\ln(u)}{(1-u)}
\biggl]\, . 
\ee

\section{\normalsize \bf Discussion}

Recall that due to the positive charge parity, the pion pair is produced in a
configuration symmetric with respect to interchange of two pions.  Obviously,
the production amplitude \re{Tdecompos} should exhibit the same symmetry.
As the part involving $T_1$ is proportional to $k_\perp$, which is odd under
exchange of pions, $T_1$ itself must be an odd function of $\xi$ as well.
Note that although $H^Q(u,\xi)$ and $\Phi^Q(u,\xi)$ are symmetric functions of
$\xi$, the presence of $\xi$-derivative in expressions \re{T1real} and
\re{T1fin} indeed results in $T_1$ which is antisymmetric function of $\xi$.

In order to estimate the magnitude of $T_1$, we recall the 'minimal' model of
the quark $2\pi$ DA \cite{Diehl1,KM00}:
\be{DAmodel}
\ba{l}
\displaystyle
\Phi^Q (u, \xi, W^2)=
-30 u(1-u)\ (2u-1)\frac1{N_f} M^Q(\mu^2) B(\xi,W) \, .
\ea 
\ee
Here $M^Q(\mu^2)$ is the momentum fraction carried by quarks in a pion at a
scale $\mu^2$. Function $B(\xi,W)$ is related to the so-called Omn\`es
functions \cite{omnes} $f_0(W)$ and $f_2(W)$ by
\be{Bdef}
B(\xi,W) = \Biggl[
\frac{3C-\beta^2}{3}\ f_0(W)\ P_0(\cos\theta_{\rm cm}) -
\frac{2}{3}\ \beta^2 \ f_2(W) \ P_2(\cos\theta_{\rm cm})
\Biggr]
\, ,
\ee
where $C=1+O(m_\pi^2)$. Omn\'es functions were analyzed in details in
\cite{Gasser}. $f_0(W)$ and $f_2(W)$ can be obtained from dispersion relations
derived in \cite{Polyakov}: 
\be{fDef}
f_l(W)=
\exp\biggl[i\delta_l^0(W)+
\frac{W^{2}}{\pi}
{\rm Re} \int_{4m_\pi^2}^\infty
ds \frac{\delta_l^0(s)}{s(s-W^2-i0)}
\biggr]\, .
\label{omnes}
\ee
Here $\delta_l^0$ are $\pi\pi$ scattering phase shifts in
the correspondig channels.

After taking the derivative over $\xi$, the $f_0$ contribution
vanishes ane one finds:
\be{T1model}
T_1^{\gamma\pi\pi}(W,\xi)=- \frac{10}{3}\ \sum e_q^2 \ \frac1{N_f} M^Q(\mu^2)
f_2(W) \xi \, .
\ee
As a result, the corresponding contribution to the production amplitude
\re{Tdecompos} has in the CM system the angular dependence $\sim
\sin(\theta_{\rm cm})\cos(\theta_{\rm cm}) e^{\pm i \phi_{\rm cm}} \sim
Y^l_m(\theta_{\rm cm},\phi_{\rm cm})$ with $l = 2$ and $m = \pm 1$. As far as
the $W$ dependence is considered, the Omn\'es function $f_2(W)$ is small above
the threshold, where pions are produced mostly in the S-wave \cite{KMP}. Due to
$f_2(1270)$ resonance, $f_2(W)$ has a peak at $W = 1.275$ GeV. This range of
$W$ is therefore most suitable to study $T_1$ experimentally, and the
$W$-dependence provides an important test of the formula (\ref{T1model}).

As pointed out in \cite{Diehl1}, by applying different weighting and averaging
procedures one can directly compare the QCD predictions for different helicity
amplitudes with the $\gamma^* \gamma \to \pi \pi$ data. The analysis of $T_1$
presented here allows to answer the question about the magnitude of the matrix
elements of quark-gluon operators, which are not accounted for in the WW
approximation. By comparing prediction for the helicity-zero amplitude $T_0$
with the data one can constraint a model of $2\pi$ quark distribution
amplitude. The same model can be subsequently used to predict the amplitude
$T_1$ according to (\ref{T1real}) and (\ref{T1fin}). Any 
discrepancy between such a prediction and the data would hint at a very
interesting scenario where a significant contribution to $T_1$ arises from
twist-3 quark-gluon operators.\\

{\bf Acknowledgments}: 
We acknowledge useful discussions with V. Braun, M.V. Polyakov and O. Teryaev. 
This work was supported in part by
KBN grant 2~P03B~011~19 and DFG project No. 920585.

\newpage

\begin{figure}[p]
\unitlength1mm
\begin{center}
\hspace{0cm}
\insertfig{8}{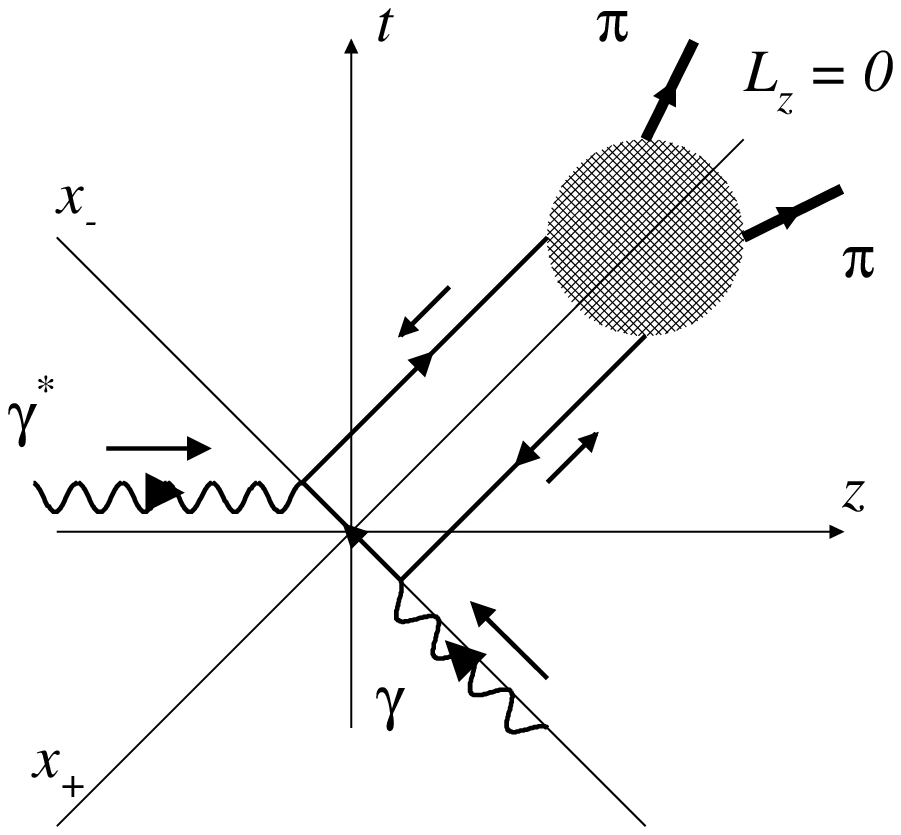}
\end{center}
\vspace{-0.5cm}
\caption{\small  
   Space-time development of the $\gamma^* \gamma$ collision. Arrows denote
   polarizations of photons and quarks, respectively. As the final $q {\bar q}$
   pair has no orbital angular momentum, its transition into the pion pair is
   described by a matrix element of twist-2 light-cone quark string operator.
\label{SpaceTime0}}
\end{figure}

\vskip2cm

\begin{figure}[p]
\unitlength1mm
\begin{center}
\hspace{0cm}
\insertfig{8}{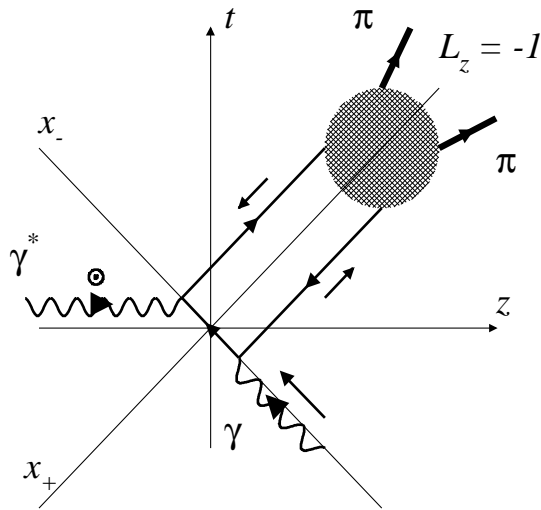}
\end{center}
\vspace{-0.5cm}
\caption{\small  
  Space-time development of the $\gamma^* \gamma$ collision. Arrows denote
  polarizations of photons and quarks, respectively. The final $q {\bar q}$
  pair carries a unit of orbital angular momentum. Its transition into the pion
  pair is described by a matrix element of a twist-3 operator.
\label{SpaceTime1}}
\end{figure}

\newpage

\begin{figure}[p]
\unitlength1mm
\begin{center}
\hspace{0cm}
\insertfig{8}{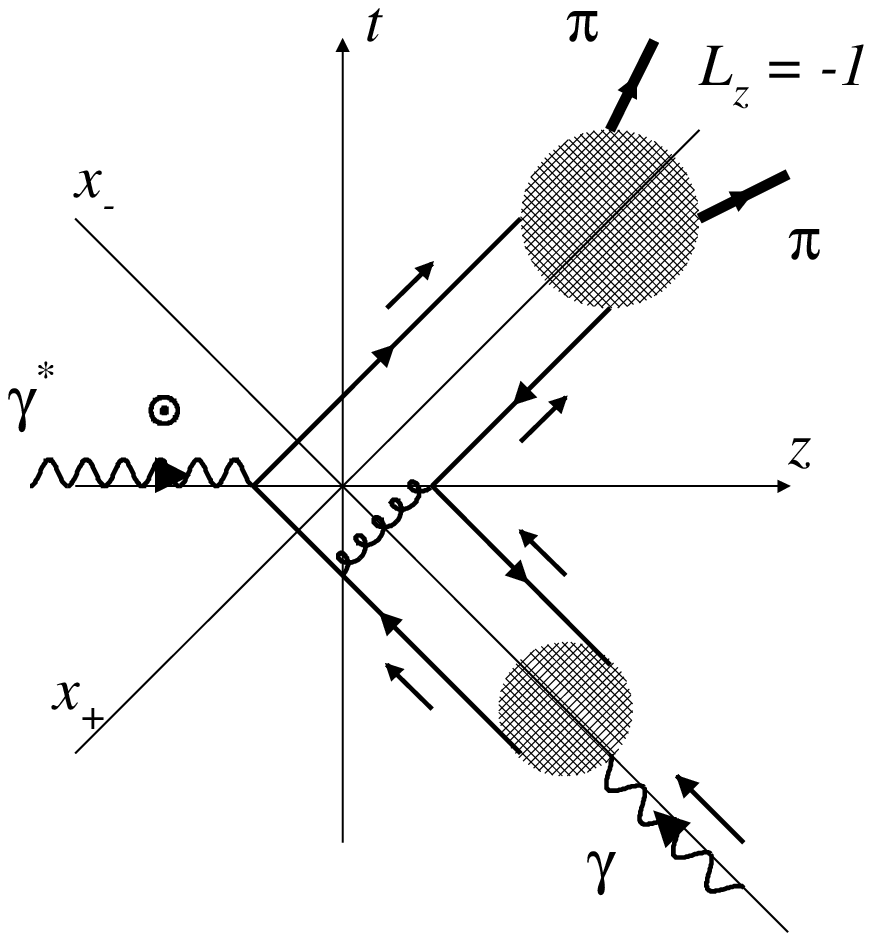}
\end{center}
\vspace{-0.5cm}
\caption{\small  
  The real photon splits into a quark-antiquark pair long before a collision
  with the virtual one. Arrows denote polarizations of photons and quarks,
  respectively. The final $q {\bar q}$ pair has to carry two units of orbital
  angular momentum and therefore its transition into the pion pair must be
  described by a matrix element of a higher-twist operator.
\label{SpaceTime2}}
\end{figure}


\begin{thebibliography}{99}

\bibitem{CELLO}
H.~J.~Behrend {\it et al.}  [CELLO Collaboration],
Z.\ Phys.\  {\bf C49} (1991) 401.

\bibitem{CLEO}
J.~Gronberg {\it et al.}  [CLEO Collaboration],
Phys.\ Rev.\  {\bf D57} (1998) 33
[hep-ex/9707031].

\bibitem{Budnev}
V.~M.~Budnev, I.~F.~Ginzburg, G.~V.~Meledin and V.~G.~Serbo,
Phys.\ Rept.\  {\bf 15} (1974) 181.

\bibitem{Lepage}
G.~P.~Lepage and S.~J.~Brodsky,
Phys.\ Rev.\  {\bf D22} (1980) 2157.

\bibitem{Efremov}
A.~V.~Efremov and A.~V.~Radyushkin,
Theor.\ Math.\ Phys.\  {\bf 42} (1980) 97.

\bibitem{Dittes}
D.~Muller, D.~Robaschik, B.~Geyer, F.~M.~Dittes and J.~Horejsi,
Fortsch.\ Phys.\  {\bf 42}, (1994) 101

\bibitem{Diehl}
M.~Diehl, T.~Gousset, B.~Pire and O.~Teryaev,
Phys.\ Rev.\ Lett.\  {\bf 81} (1998) 1782
[hep-ph/9805380].

\bibitem{Freund}
A.~Freund,
Phys.\ Rev.\  {\bf D61} (2000) 074010
[hep-ph/9903489].

\bibitem{Polyakov}
M.~V.~Polyakov,
Nucl.\ Phys.\  {\bf B555} (1999) 231
[hep-ph/9809483].

\bibitem{Ji97}
X. Ji, \PRD{55}{7114}{1997}.

\bibitem{Rad97} A.V. Radyushkin,
\PRD{56}{5524}{1997}.

\bibitem{KMP} N.Kivel,L.Mankiewicz, M.V.Polyakov,
Phys. Lett. B {\bf 467}, (1999), 263.

\bibitem{Diehl1}
M.~Diehl, T.~Gousset and B.~Pire,
hep-ph/0003233.

\bibitem{Wandzura}
S.~Wandzura and F.~Wilczek, \PLB{72}{1977}{195}.

\bibitem{g2}
P.~Bosted et.all., \NPA{663}{2000}{297}.

\bibitem{instanton}
J.~Balla, M.~V.~Polyakov and C.~Weiss, \NPB{510}{1998}{327}.\\
B.~Dressler, M.~V.~Polyakov, \PRD{61}{2000}{097501}.

\bibitem{Anikin}
I.~V.~Anikin, B.~Pire and O.~V.~Teryaev,
\PRD{62}{2000}{071501}.

\bibitem{Penttinen}
M.~Penttinen, M.~V.~Polyakov, A.~G.~Shuvaev and M.~Strikman,
``DVCS amplitude in the parton model,''
hep-ph/0006321.

\bibitem{BM}
A.~V.~Belitsky and D.~Muller,
``Twist-three effects in two-photon processes,''
hep-ph/0007031.

\bibitem{KPST}
N.~Kivel, M.~V.~Polyakov, A.~Schafer and O.~V.~Teryaev,
``On the Wandzura-Wilczek approximation for the twist-3 DVCS  amplitude,''
hep-ph/0007315.

\bibitem{RadWeiss}
A.~V.~Radyushkin,~C.~Weiss, `DVCS amplitude with kinematical twist-3 terms,''
hep-ph/0008214. 

\bibitem{BlRo}
J.~Blumlein and D.~Robaschik,
\NPB{581}{2000}{449}.

\bibitem{Rad}A.~V.~Radyushkin,
Phys.\ Rev.\  {\bf D56} (1997) 5524
[hep-ph/9704207].

\bibitem{PW}
M.~V.~Polyakov and C.~Weiss,
Phys.\ Rev.\  {\bf D60} (1999) 114017.

\bibitem{photon}
B.~L.~Ioffe, A.~V.~Smilga,
\NPB{232}{1984}{109}.

\bibitem{BB89}
I.~I.~Balitsky and V.~M.~Braun,
Nucl.\ Phys.\  {\bf B311} (1989) 541.

\bibitem{MPW}
L.~Mankiewicz, G.~Piller, T.~Weigl,
\EPJC{5}{1998}{119}. 

\bibitem{CollinsDiehl}
J.~C.~Collins, M.~Diehl (DESY).
\PRD{61}{2000}{114015}.

\bibitem{KM00}
N.~Kivel, L. Mankiewicz,``Power Corrections to the process $\gamma^* \gamma \to
\pi \pi$ in the Light-Cone Sum Rules Approach´´,hep-ph/0008168. 

\bibitem{omnes}
R.~Omn\`es, Nuovo Cim. {\bf 8}, 316 (1958).

\bibitem{Gasser} J.F. Donoghue, J.~Gasser and H.~Leutwyler,
Nucl. Phys. B {\bf 343}, 341 (1990).






 






\end{thebibliography}
\end{document}